\documentstyle[prl,aps,multicol,epsf]{revtex}

\def\mathrm#1{\mbox{#1}}
\def\approxeq{\approx}
\newcommand{\be}{\begin{equation}}
\newcommand{\ee}{\end{equation}}
\newcommand{\bea}{\begin{eqnarray}}
\newcommand{\eea}{\end{eqnarray}}

\newcommand{\htc}{high-T$_c$ }

\newcommand{\cacocl}{Ca$_{2}$CuO$_{2}$Cl$_{2}$ }
\newcommand{\tj}{$t$-$J$ }
\newcommand{\tp}{t^\prime}
\newcommand{\tpp}{t^{\prime\prime}}
\newcommand{\INT}{int}
\newcommand{\kin}{kin}
\newcommand{\neel}{N\'{e}el }
\newcommand{\eps}{\varepsilon}
\newcommand{\af}{AF}
\newcommand{\SC}{SC}
\newcommand{\mod}{mod}
\newcommand{\sgn}{\mathrm{sgn}}
\newcommand{\up}{\uparrow}
\newcommand{\down}{\downarrow}
\newcommand{\ul}{}
\begin{document}
\draft

\title{Interrelation of Superconducting and Antiferromagnetic Gaps in \htc
Compounds: a Test Case for the SO(5) Theory}

\author{Marc G.\ Zacher$^1$, Werner Hanke$^{1,2}$, Enrico Arrigoni$^1$, 
Shou-Cheng Zhang$^2$}
\address{$^1$Institute for Theoretical Physics, University of  W\"urzburg,
97074 W\"urzburg, Germany\\
$^2$ Department of Physics, Stanford University, Stanford, California 94305, 
USA}

\maketitle

\begin{abstract}
Recent angle resolved photoemission data, which found evidence 
for a $d$-wave-like modulation of the
antiferromagnetic gap, suggest an intimate interrelation between the 
antiferromagnetic insulator and
the superconductor with its $d$-wave gap. 
It is shown here that a projected SO(5) theory, which
explicitly takes the Mott-Hubbard gap into account, 
correctly describes the observed gap characteristics.
Specifically, it
accounts for the order of magnitude difference
between the antiferromagnetic gap modulation and the superconducting gap and
is also consistent with the gap dispersion.
\end{abstract}
\pacs{PACS numbers:
74.20.-z, 
11.30.Ly, 
74.25.Jb  
79.60.Bm
}

\begin{multicols}{2}

\begin{figure}
\epsfxsize=8cm
\epsffile{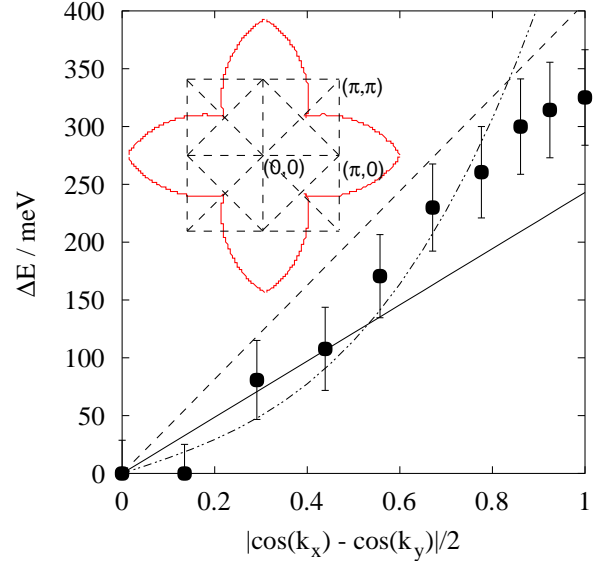}
\caption{ARPES results for Ca$_2$CuO$_2$Cl$_2$: 
$\Delta E=E(k)-E(\frac{\pi}{2},\frac{\pi}{2})$
is plotted (dots with errorbars)
along the magnetic zone boundary against the n.n. $d$-wave function.
The straight lines show the results of the SDW (full line)
and Slave Boson (dashed) calculation based on 
a n.n. 
$d$-wave dispersion $d_{\vec k}=\cos k_x - \cos k_y$ for the SC
gap (see text).
The curved dash-dotted 
line gives the SB result using the extended
$d$-wave function obtained from the fit of the Bi2212 SC gap
in Fig. \protect\ref{figmesot}.  
}
\label{figshen}
\end{figure}

Angular-resolved photoemission spectroscopy (ARPES) recently provided
evidence which points to a 
direct correlation between the $d$-wave symmetry of the
superconducting (SC) gap and an observed $d$-wave-like modulation of the 
antiferromagnetic (AF) gap \cite{ronning}.
The ''gap'' structure in the AF phase, as found by the ARPES experiments in
insulating \cacocl, is summarized in Fig. \ref{figshen}. These
data display a $d$-wave-like
dispersion in the one-electron spectral function $A(\vec k ,\omega)$ 
with respect to the lowest energy state at $(\frac{\pi }{2},\frac{\pi }{2}) $,
as revealed in the inset of this Figure. 
This becomes also obvious when the energy difference $E(\vec k) - 
E(\frac{\pi }{2},\frac{\pi}{2})$ is plotted versus the simple 
nearest-neighbor (n.n.) $d$-wave dispersion
 $|\cos k_x - \cos k_y|$.
This ``$|d|$-wave'' like
gap \cite{notedisp} is a modulation of the uniform ($s$-wave) Mott-Hubbard 
gap of the order of $U\sim eV$ in the insulating state. 
The crucial point is that
these photoemission data suggest that the $|d|$ component of the AF gap in the
insulator is also the underlying reason for the celebrated pseudo-gap in the
underdoped regime: this ''high-energy'' pseudo-gap of the order 
of the magnetic exchange $J\sim 0.2eV$
continuously evolves out of the insulating feature, as documented not only
by the same energy scale but again by the same $d$-wave dispersion 
\cite{marshall}.
Since, on the other hand, this high-energy feature is
closely correlated to the SC gap 
as a function of both doping and momentum 
\cite{marshall,ding,white},
we finally arrive at a constraint on the
microscopic theory: such a theory should be able to explain the
interrelation between the SC gap and the AF gap modulation.

In this Letter, it is shown that 
a modified
version of the SO(5) theory of \htc superconductivity, i.e. the
{\it projected} SO(5) theory,
provides rather naturally such an interrelation.
This theory aims at unifying AF and SC via a
symmetry principle, while at the same time explicitly taking the
Mott-Hubbard gap into account. The projection is the crucial new ingredient, 
since the exact SO(5) symmetry requires 
charge excitations at
half-filling to have the same gap as the collective spin-wave excitations
\cite{zhang,rabello,henley}. This condition is
violated in a Mott-Hubbard insulator, which has a large gap ($\sim eV$) to 
all charge
excitations while the spin excitations display no gap. 
In particular, in an exactly SO(5) symmetric description \cite{rabello,henley}, the
SC gap with its nodes would directly be mapped onto an AF gap,
 which then would
have precisely the same magnitude and would vanish at the nodes 
$(\pm \frac{\pi}{2},\pm \frac{\pi}{2})$. 
Taking the Mott--Hubbard gap into account amounts to properly projecting out
the ''high-energy'' charge
processes of order $\sim eV$ (Gutzwiller constraint) 
from the ''low-energy'' SO(5) rotation between AF and SC states \cite{zhanghu}.

Our two main results are:

(i) 
The projected SO(5) symmetry naturally 
introduces the $s$ component of the AF gap
associated with the large on-site Coulomb energy, which is absent in the case of
pure SO(5) symmetry, and quantitatively relates the remaining $|d|$ component of the
AF gap with the $d$-wave SC gap:
In accordance with experiment,
the $d$-wave SC gap is found to be of the order of $J/10$ whereas
the obtained AF gap modulation is of the order of $J$.
This is a direct consequence of the projection.

(ii) 
The projected SO(5) theory accounts also for the dispersion of the two gap 
structures. It can also include deviations from the
simple $\cos k_x - \cos k_y$ form,
which have been recently reported both for SC \cite{mesot} and AF 
\cite{shenpriv} gaps.

Besides finding a modulation of the gap in the AF insulator, the recent
ARPES experiment\cite{ronning} also found a remnant Fermi surface of the
AF insulator. This shows that it is appropriate to think of the AF insulator
in terms of a condensate of {\it magnons} on top of a Fermi-liquid like state,
just like a superconductor can be viewed as a condensate of {\it Cooper pairs}
on top of a Fermi-liquid state.
To be more specific, the variational 
wave function of an AF insulator \ul{with \neel vector pointing in 
$\alpha$--direction}
is given by 
\begin{equation}
|\Psi_{\af} >\sim \prod_k (u_k + v_k c^\dagger_{k+Q} 
\sigma_\alpha c_k) |\Omega\rangle ,
\label{af}
\end{equation}
where $c^\dagger_k$ is the spinor creation operator 
with wave vector $k$, $Q=(\pi,\pi)$ is the AF vector, $\sigma_\alpha$
are the Pauli spin matrices, $u_k$ and $v_k$ the variational
parameters and $|\Omega\rangle$ is a half-filled Fermi-liquid like
state.
A magnon is defined by the triplet operator $N_\alpha(k)=c^\dagger_{k+Q} 
\sigma_\alpha c_k$.
On the other hand, a SC state is described by a formally completely equivalent 
state, where $u_k$ and $v_k$ are now the usual variational parameters, and 
the Cooper pair operator $B(k)=c_k \sigma_y c_{-k}$ replaces the magnon 
operator $N_\alpha(k)$ in Eq. (\ref{af}).
This ``replacement" is exactly provided by the SO(5) rotation operator
\begin{equation}
\pi_\alpha = \sum_k g_k c_{k+Q} \sigma_\alpha \sigma_y c_{-k} \, , 
\label{pi}
\end{equation}
where the form factor  $g_k$
can be written as
$g_{\vec k} = \sgn( d_{\vec k} )$. 
Here, $d_{\vec k}$
is the dispersion of the $d$-wave SC gap
and for n.n.
 pairing would be given by the simple $\cos k_x - \cos k_y$
form. 
Recent experiments by Mesot {\it et al.} indicate a deviation from this
simple n.n.\ expression. This can be taken into account by choosing
$d_{\vec k}= b (\cos k_x - \cos k_y) + (1-b) (\cos 3k_x - \cos 3k_y)$\cite{noteb2b3}.
The parameter $b$ (for $b\neq 1$) emphasizes the
importance of longer-ranged (3rd n.n.) pairings \cite{formfactor}.

The $\pi$-operator rotates the magnon and Cooper pair operators into each other
according to the following equation:
\begin{equation}
[\pi_\alpha, N_\beta(k)] = \delta_{\alpha\beta} g_k B(k) \ \ , \ \
[\pi^\dagger_\alpha, B(k)] = g_k N_\alpha(k) \, .
\label{rotation}
\end{equation}
From
the above equation we see that,
within an SO(5) symmetric description,
a $d$-wave
form 
of the Cooper pair wave function will translate quite generally into 
a $|d|$-form  of the magnon wave function.

We want now to explore more quantitatively the consequences of
assuming such a projected SO(5) symmetry for the high-T$_c$ cuprates.
Our starting point is the 
SC state.
We first choose
the simplest fermionic lattice hamiltonian which reproduces the
$d$-wave SC state of the \htc materials in a 
simple BCS mean-field description.
 We then perform an SO(5) rotation on 
the operator-level that introduces the magnetic part of the 
interaction.
 The resulting SO(5)-invariant hamiltonian takes thus the form
\begin{eqnarray}
H_{\kin}+H_{\INT} &=& 
\sum_{p,\sigma }\varepsilon _{p}c_{p,\sigma
}^{\dagger}c_{p,\sigma }+
\frac{V}{N}\sum_{\vec r_1,\vec r_2} \Big\{ 
- \vec m(\vec r_1)\cdot 
\vec m(\vec r_2) 
\nonumber \\
& & + \frac{1}{2}\left( \Delta(\vec r_1)\Delta^{\dagger}(\vec r_2)+
\Delta^{\dagger}(\vec r_1)\Delta(\vec r_2)\right) \Big\} .  \label{gl2}
\end{eqnarray}

Here, $H_{\kin}$ stands for the kinetic energy part with band dispersion 
$\varepsilon _{p}=-2t\left( \cos p_x+\cos p_y\right) $, valid for a
nearest-neighbor tight-binding model with hopping amplitude $t$. 
$H_{\INT}$ contains a spin-spin interaction and a
pair-hopping term \cite{rabello,henley}.
The SC part of $H_{\INT}$ is of reduced BCS form and is given in momentum space
by $2 V \{ \Delta \Delta^\dagger +
\Delta^\dagger \Delta \}$, where $\Delta$ is the usual 
$d$-wave order parameter
$\Delta = \sum_{p} \frac{1}{2} d_{\vec p} c_{\vec p,\up}
c_{-\vec p,\down}^{}$. This 
BCS form in $\Delta$ and $\Delta^\dagger$ fixes the
general SO(5)-invariant interaction with the coupling 
$V(\vec p, \vec p^{\, \prime}; \vec q)$ to
be separable in momentum space and given by 
$V(\vec p, \vec p^{\, \prime}; \vec q) = 
V \delta_{\vec q,\vec Q} |d_{\vec p}| |d_{\vec p^{\, \prime}}|$.
On the other hand, the \neel order parameter $\vec m(\vec r)$ has an
 extended
internal structure \cite{noteintstruc}.
This  internal structure 
is required by the SO(5)
symmetry, and may, at least in principle, be tested in experiments. 
In particular, on a mean-field level, it may be related effectively 
to spatially extended hoppings $\tp$ and $\tpp$ \cite{noteintstruc}, 
which have previously been introduced as parameters 
in \tj and Hubbard models
to account for the AF gap anisotropy 
\cite{eder97,duffy,kim}.

The Gutzwiller projection, which reduces the full SO(5) symmetry to a
{\it projected} SO(5) symmetry, can be implemented by the introduction
of a Hubbard $U$ interaction and by taking the limit of large $U$. 
Therefore, we arrive at the following hamiltonian 
\begin{equation}
H=\left( H_{\kin}+H_{\INT}\right) +H_{U}+H_{\mu },  \label{g3}
\end{equation}
where 
$H_{U}=U\sum_{i}n_{i\uparrow }n_{i\downarrow }$
is the
standard Hubbard interaction in real space
($n_{i\sigma }=c_{i\sigma }^{\dagger}c_{i\sigma }^{}$) 
and $H_{\mu }=-\mu \sum_{i,\sigma} n_{i\sigma}$ denotes the chemical
potential term.

The hamiltonian (\ref{g3}) is further motivated by recent numerical
and analytical results on 
much-used
 hamiltonians \cite{meixner,ederetal98,arri}, in particular the \tj and
 Hubbard model.
These results have shown the presence of an approximate SO(5) symmetry 
in the low-energy bosonic excitations. 
However, the \tj model cannot explain the $|d|$ AF gap modulation in the fermionic sector.
It, therefore, misses an important piece of physics our current model contains (unless one
introduces {\it ad-hoc} values for $t^\prime$, $t^{\prime\prime}$, see discussion above).
The logic of our approach is similar in spirit to the phenomenological "Landau approach" to
strongly correlated systems; i.e. rather then starting from first principles we construct an
effective model from simple symmetry principles, and check whether it reproduces the low
energy experiments.

There are various ways to  study the hamiltonian in Eq. (\ref{g3}).
Its physical content becomes transparent already on the simplest, i.e. 
Hartree-Fock mean-field level. Earlier work by Schrieffer {\it et al.} 
on the Hubbard model \cite{schrieffer} shows that such a simple
mean field calculation can capture the basic physics also in the 
strong-coupling limit. 
Consider first the Spin-Density-Wave (SDW)-type of solution for 
the \neel state. Here, the gap
function $\Delta({\vec p})$ is connected to the SDW mean-field (polarized in
$z$-direction) by the standard relation:
\begin{equation}
\langle c_{\vec p + \vec Q}^\dagger \sigma_{z} c_{\vec p}^{} \rangle 
= \frac{\Delta(\vec p)}{2 E(\vec p)} \, ,
\label{glorderparam}
\end{equation}
where, as usual, $E(\vec p) = \left( \eps^2(\vec p) + 
\Delta^2(\vec p) \right)^{1/2}$. When
introduced in Eq. (\ref{g3}) for the hamiltonian, this mean-field order 
parameter
results in the self-consistency condition determining the gap $\Delta(\vec p)$,
\begin{equation}
\frac{1}{N}
\sum_{\vec p^{\, \prime}} V(\vec p, \vec p^{\, \prime}) \frac{\Delta(\vec p^{\,
\prime})}
{2 E(\vec p^{\, \prime})} = \Delta(\vec p) 
\label{glselfcon}
\end{equation}
($N$ being the number of lattice sites). 
Taking the factorized form of the SO(5) interaction (\ref{gl2}), and
including the Hubbard-$U$ term, we obtain
\begin{equation}
 V(\vec p, \vec p^{\, \prime}) = U + V |d_{\vec p}| |d_{\vec p^{\, \prime}}| \, .
\label{glvppprime}
\end{equation}
First, note that the factorized form of the interaction 
$V(\vec p, \vec p^{\, \prime})$ 
introduces
a  separable form of the AF gap, 
\begin{equation}
\Delta^{\af}(\vec p) = \Delta_U + \Delta_{\mod} |d_{\vec p}| \, .
\label{glgapmodul}
\end{equation}
For large values of the Hubbard
interaction, $\Delta_U$ is of the order of $U$. 
Eq. (\ref{glgapmodul})
then establishes the gap-modulation $\sim \Delta_{mod} |d_p|$ on top of a uniform gap
in the AF.

\begin{figure}
\epsfxsize=8cm
\epsffile{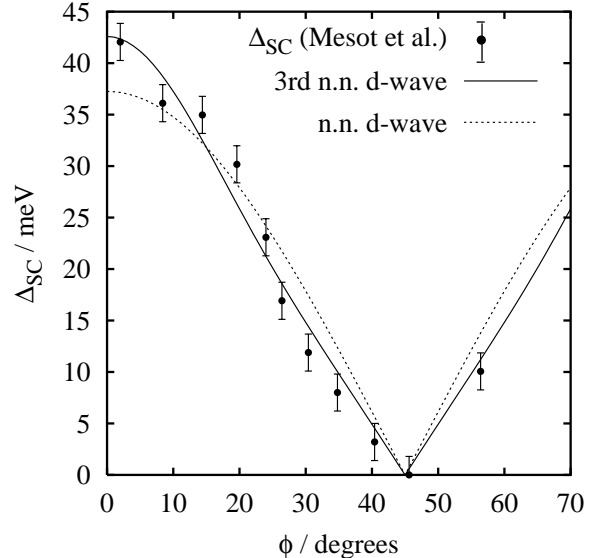}
\caption{ARPES results for the SC gap of Bi2212 in 
the underdoped region
($T_c$=75 K) taken from reference \protect\cite{mesot} plotted 
against the fermi-angle $\phi$
(45 degrees corresponds to $k_x = k_y$ and 0 degrees to maximal $k_x$).
The lines show fits to the data with both
n.n. $d$-wave and 3rd n.n. $d$-wave functions
($d_{\vec k}= b (\cos k_x - \cos k_y) + (1-b) (\cos 3k_x - \cos 3k_y)$). 
With this fit procedure to the SC state, all free
parameters in the theory are fixed (taking $t = 0.5 eV$ as the energy scale).
}
\label{figmesot}
\end{figure}

Second, in formal analogy, in the $d$-wave SC state, the same gap equation 
(\ref{glselfcon})
holds.
As discussed in Eq. \ref{rotation}, the $|d_{\vec p}|$-form factor has to 
be replaced by 
$|d_{\vec p}| g_{\vec p}= d_{\vec p}$ and, therefore,
the relevant interaction is 
$V(\vec p, \vec p^{\, \prime}) = V \, d_{\vec p} d_{\vec p^{\, \prime}}$
(the $U$-term drops out) resulting in the gap function:
\begin{equation}
\Delta^{\SC}(\vec p) = \Delta_{\SC} \cdot d_{\vec p} \, .
\label{glgapsc}
\end{equation}
Thus, both the AF gap in equation (\ref{glgapmodul}) and the SC gap have 
the required form.
Our strategy now is to fit quantitatively the SC experimental gap (Fig.
\ref{figmesot}), then perform the SO(5) rotation, and compare the so 
obtained AF gap
features with the ARPES results in Fig. \ref{figshen}.
The SC gap is fixed in
accordance with new ARPES data \cite{mesot}, allowing also for longer-ranged 
(3rd n.n.)
interactions, obtained by using $b\neq 1$ ($b=0.81$) in the extended $d$-wave
form $d_{\vec k}$.
However, we expect the precise value of $b$ to be model dependent and
to differ for the two materials studied in Refs. \cite{ronning} and \cite{mesot}.
The SO(5)-coupling strength $V$ in Eq. (\ref{gl2}) 
was chosen in such a way that 
(for both BCS and Slave-Boson (SB) evaluations, see below) it
gives a $d$-wave gap of the correct order of magnitude 
in the SC phase, i.e. $\Delta_{\SC}=0.04t\approxeq J/10$ ($t=0.5 eV$)
\cite{notespecificmeanfield}.

Experiments tell us
 that while $\Delta_{\SC}$ is of the order $J/10$,
$\Delta_{\mod}$ is an order of magnitude larger. 
To verify this independently from the above SDW/BCS evaluation,
we have additionally used the 
Slave-Boson formalism, 
which we treat by the usual saddle-point approximation 
\cite{kotliar,arrigonizacher}.
The essential observation, independent from the specific mean-field treatment, 
is then that, while the SO(5) interaction is responsible for the 
$d$-wave structure of both gaps,
it is a different mechanism, namely the Hubbard gap, which is
responsible for the experimentally observed order of magnitude differences
in $\Delta_{\SC}$ and 
$\Delta_{\mod}$.

\begin{figure}
\epsfxsize=8cm
\epsffile{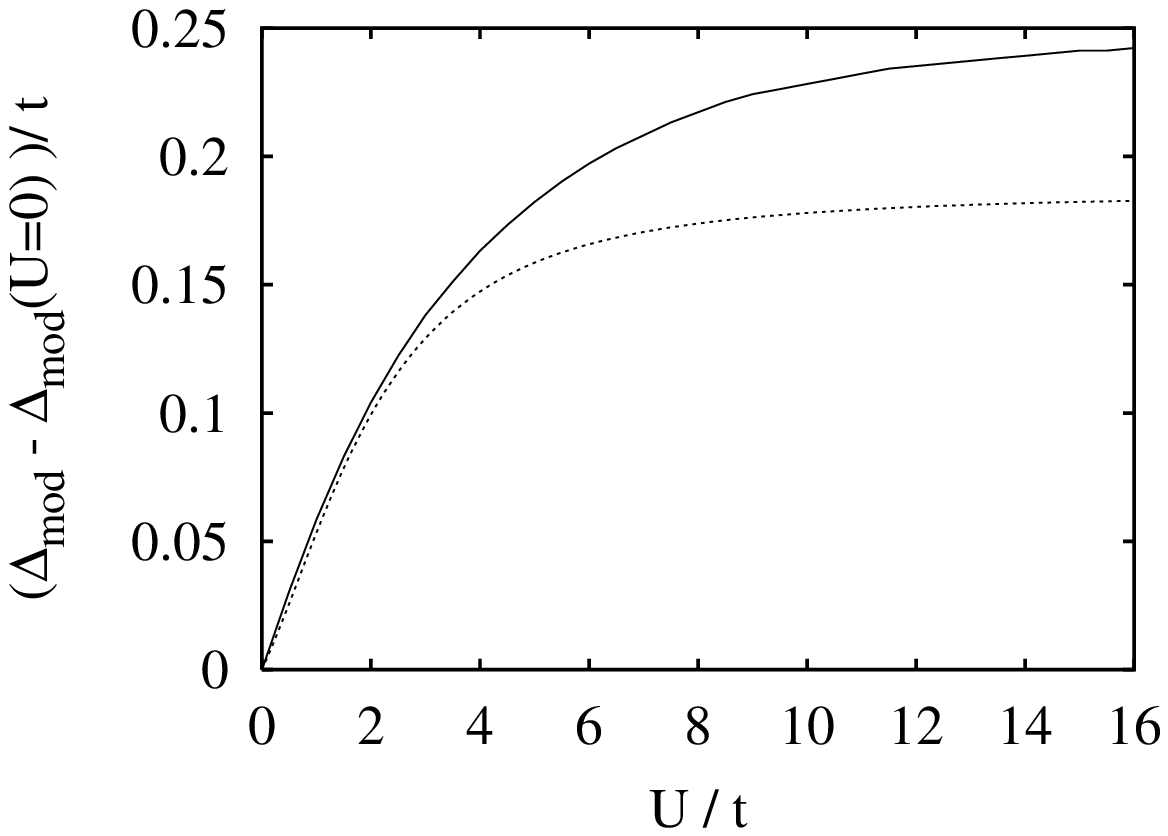}
\caption{
The AF gap modulation $\Delta_{\mod}-\Delta_{\mod}(U=0)$
{\protect \cite{deltamodu0}}
 obtained from the SDW (dotted line)
and Slave Boson (full line) calculation as a
 function of the
symmetry-breaking Hubbard interaction $U$. 
Note that the $\Delta_{\mod}$-value for $U=8t$ is already close to the 
saturated value
obtained for complete projection ($U\rightarrow \infty$).
}
\label{figgapu}
\end{figure}

This is clearly demonstrated in Fig. \ref{figgapu}, which plots the amplitude 
$\Delta_{\mod}$ of the AF $d$-wave-like modulation as a function of the
Hubbard interaction $U$. We
note that 
increasing $U$  and, therefore,
suppressing the doubly occupied states, strongly enhances the
 $\Delta_{\mod}$
in both the SDW and SB evaluations.
It is comforting to notice that the results are already converged at
the commonly accepted value for $U=8t$, i.e. the projection is almost 
complete here.
Taking this value of $U$
yields
an AF gap modulation $\Delta_{\mod}\approxeq
0.24\ t$ for SDW, and $\approxeq 0.41\ t$ for SB. Thus,
we find a radically different energy scale between $\Delta_{\mod}$ 
($\sim J$) and $\Delta_{\SC}$ ($\sim J/10$), in agreement
with the ARPES data.

To summarize, our results on the $d$-wave dispersion, $d_{\vec k}$ 
in the 
SC gap and $|d_{\vec k}|$ in the AF gap modulation, are in general
accordance with recent ARPES data \cite{mesot,ronning}. Thus,
the {\it projected} SO(5) rotation provides a definite link between the data
points observed in 
two
quite different phases, i.e. the insulating AF,
and the SC.
This concept of projection is crucial, since, if one  had used
an exact SO(5)
theory, without the physically relevant term $H_{U}$, one would have obtained
an AF gap with nodes.
Moreover, as demonstrated here, $H_{U}$ is
pivotal in explaining the order of magnitude differences 
between the SC gap $\Delta_{\SC}$ and the $d$-wave-like 
modulation of the AF gap $\Delta_{\mod}$.
Just like the neutron resonance mode can be interpreted
as the reflection of AF correlation in the SC state
\cite{zhang,meixner}, the ARPES experiment
can be interpreted as the reflection of the SC correlation in the AF state.

We thank Z.-X.\ Shen, R.B.\ Laughlin, D.J.\ Scalapino, R.\ Eder, 
J.\ C.\ Campuzano and O.K.\ Andersen for helpful
discussions and suggestions.
W.H. and E.A.
would also like to acknowledge the support and hospitality of the 
Stanford Physics
Department, where part of this work was carried out. This work was  supported
by the DFG (AR 324/1-1 and HA 1537/17-1), DFN (Contract No. TK598-VA/D3),
BMBF (05SB8WWA1), and NSF (Grant No. DMR-9814289).

\narrowtext

\newpage

\end{multicols}


\begin{thebibliography}{99}
\bibitem{notedisp}
Notice that the ARPES
data measure only the dispersion of the first ionization states, i.e. the
''topmost'' valence band dispersion. In case of electron-hole symmetry, as
in our theory, this dispersion is then translated into an equivalent 
$|d_{\vec k}|$-modulation of the AF gap.
\bibitem{noteb2b3}
In Ref. {\protect \cite{mesot}}, an extension of the
nearest neighbor $d$-wave of the form $\cos 2k_{x/y}$ 
was chosen whereas we
use a $\cos 3k_{x/y}$ form. Both describtions fit the 
experimental data points equally well 
(see Fig. 2 in Ref. {\protect \cite{mesot}}
and our Fig. {\protect\ref{figmesot}}).
\bibitem{formfactor}
Notice that this generalized form factor $g_{\vec k}$ satisfies the
SO(5) group closure
requirements in the same way as the
$\sgn(\cos k_x - \cos k_y)$ form introduced in 
Refs.~\cite{rabello,henley}. Our form has the advantage of a more
appropriate description of the experiments.
\bibitem{noteintstruc}
The extended internal structure of $\vec m(\vec r_1)$ 
[i.e. $\vec m(\vec r_1) = \sum_{\vec
r_2} \frac{1}{2} w(\vec r_1 - \vec r_2) e^{i \vec Q \vec r_2} c^\dagger_{\vec r_1} \vec
\sigma c^{}_{\vec r_2}$, where $\vec \sigma$ are the usual $2 \times 2$ Pauli
matrices] introduces on a mean field level (replacing $\vec m(\vec r_2)$ in Eq.
(\ref{gl2}) by twice its expectation value $\langle \vec m(\vec r_2) \rangle$) 
longer-ranged
hopping terms $\tp$ (next n.n.) and $\tpp$ (third n.n.). Similarly to the 
commonly used 
spatially extended hoppings \cite{kim,eder97,duffy}, their signs alternate. 
This stems from the fact that $w(\vec r)$ is the Fourier-transform of 
$|d(\vec p)|$ 
and, for n.n. $d$-wave,
takes the values ($\vec r = (m,n)$ on the 2D
lattice): $w(m,n)=0$, if $m+n=$ odd, $w(0,0)=\frac{8}{\pi^2}$, 
$w(1,1)=-\frac{8}{3\pi^2}$
and $w(2,0) = \frac{8}{9 \pi^2}$, etc. 
This results in $\tp = \Delta_{\mod} w(1,1) \approx 0.07 t$ 
and $\tpp = \Delta_{\mod} w(2,0)$ with $\tpp/\tp = - 1/3$, where we 
used $\Delta_{\mod}=0.243 t$ 
(the result of our n.n. SDW calculation).
\bibitem{notespecificmeanfield}
For both the SB and SDW/BCS mean-field evaluations with and without 
3rd-n.n.-pairing 
the starting point was to fit the SC gap to the Mesot experiment. 
At the fixed doping of 11 \% 
this results in $V = 0.61 t$ (n.n. SDW/BCS),
$V=0.61 t$ (n.n. SB), $V=0.96 t$ (3rd n.n. SDW/BCS), $V=1.66 t$ (3rd n.n. SB).
\bibitem{deltamodu0}
Note that we plot in Fig. \ref{figgapu} the difference
$\Delta_{\mod} - \Delta_{\mod}(U=0)$ since $\Delta_{\mod}(U=0)$ differs
for SDW and SB due to the slightly different $V$-value resulting from the
fit to $\Delta_{\SC}$.
\end{thebibliography}
\end{document}